\documentclass{PoS}

\title{Muoproduction of exotic charmonia at COMPASS}

\ShortTitle{Muoproduction of exotic charmonia at COMPASS}

\author{\speaker{A. Guskov}\thanks{On behalf oh the COMPASS Collaboration}\\
        Joint Institute for Nuclear Research, Dubna, Russia\\
        E-mail: \email{avg@jinr.ru}}

\abstract{Exotic charmonium-like states have been targeted by various experiments in the last 15 years,
but their nature still is unknown. Photo-(muo)production is a new promising instrument to study them. 
COMPASS, a fixed target experiment at CERN, analyzed the full set of the data collected with a
muon beam between 2002 and 2011, covering the range from 7 GeV to 19 GeV in the centre-of-mass energy
of the (virtual)photon-nucleon system. Production of the X(3872) state in the reaction $\mu^+~N \rightarrow \mu^+(J/\psi\pi^+\pi^- )\pi^{\pm} N'$ has been observed with a statistical significance of around 5 $\sigma$. The shape of the $\pi^+\pi^-$ mass distribution from the decay $X(3872)\rightarrow J/\psi\pi^+\pi^-$ shows disagreement with previous observations. The product of the cross section and the branching fraction of the $X(3872)$ decay into $J/\psi\pi\pi$ is estimated as 71$\pm$28(stat)$\pm$39(syst) pb.
The results obtained for the production of the $Z_c^{\pm}(3900)$ are also reported as well as future perspectives.}

\FullConference{XVII International Conference on Hadron Spectroscopy and Structure - Hadron2017\\
		25-29 September, 2017\\
		University of Salamanca, Salamanca, Spain}

\begin{document}

\section{Introduction}
In the last years a lot of new charmonium-like hadrons, the so-called XYZ states, at masses above 3.8 GeV/$c^2$ were discovered.
Several interpretations of the new states do exist: pure quarkonia, tetraquarks, hadronic molecules, hybrid mesons with a gluon content, etc. But till now many basic parameters of these states have not been determined yet. New experimental input is required
to distinguish between the models that provide different interpretations of the nature of exotic charmonia. Search for exotic charmonium-like states in exclusive photoproduction reactions has been proposed first in \cite{photo0,photo1,photo2}.

COMPASS, a fixed target experiment at a secondary beam of SPS (CERN) \cite{proposal, COMPASS1}, has a unique possibility to contribute to XYZ-physics by investigating photo(lepto-)production of these states. The experimental data obtained for positive muons of 160~GeV/$c$ (2002-2010) or 200~GeV/$c$ momentum (2011)  scattering off solid $^6$LiD
             (2002-2004) or NH$_3$ targets (2006-2011) were used to look for exotic charmonia production. 
             The amount of available COMPASS data is equivalent to about 14 pb$^{-1}$ of the integrated luminosity, when considering a real-photon beam of about 100 GeV incident energy scattering off free nucleons.
             
\section{Exclusive photoproduction of $X(3872)$}
The exotic hadron $X(3872)$ was discovered by the Belle collaboration in 2003 \cite{x3872belle}. Its mass is 3871.69 $\pm$ 0.17 MeV/$c^2$  that is very close to the $D^0\bar{D}^{*0}$ threshold. The decay width of this state was not determined yet, only an upper limit for the natural width $\Gamma_{X(3872)}$ of about 1.2 MeV/$c^2$ (CL$=$90\%) exists. The quantum numbers $J^{PC}$ of the $X(3872)$ were determined by LHCb to be $1^{++}$ \cite{x3872LHCB1,x3872LHCB2}. Approximately equal probabilities to decay into $J/\psi3\pi$ and $J/\psi2\pi$  final states indicate large isospin symmetry breaking.

\begin{figure}
 \begin{minipage}{18pc}
   \includegraphics[width=200px]{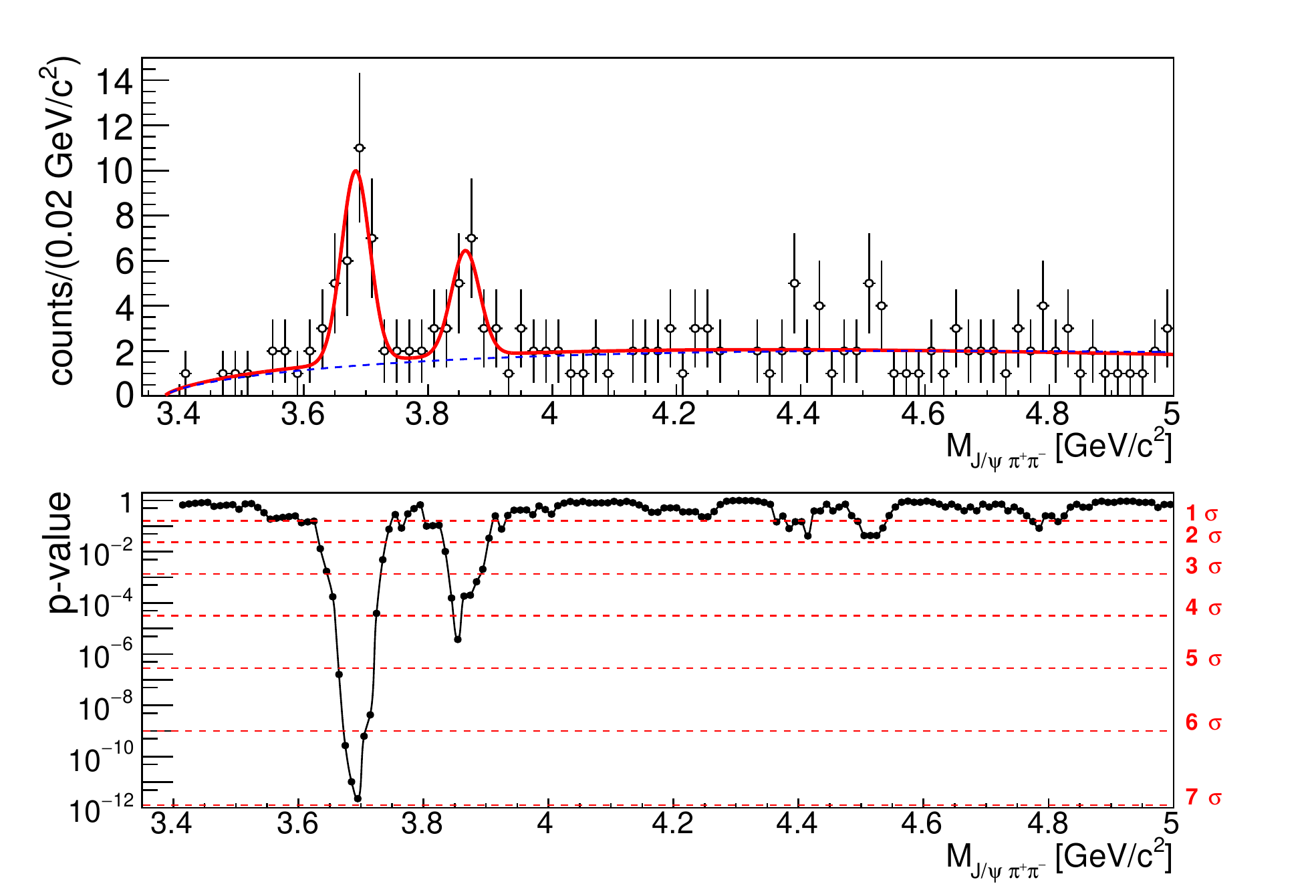}  
     \caption{\label{fig:x3872}
 The $J/\psi\pi^+\pi^-$ invariant mass distribution for the $J/\psi\pi^+\pi^-\pi^{\pm}$ final state (top).
 The statistical significance of the signal (bottom). }
\end{minipage}\hspace{2pc}%
\begin{minipage}{18pc}
   \includegraphics[width=200px]{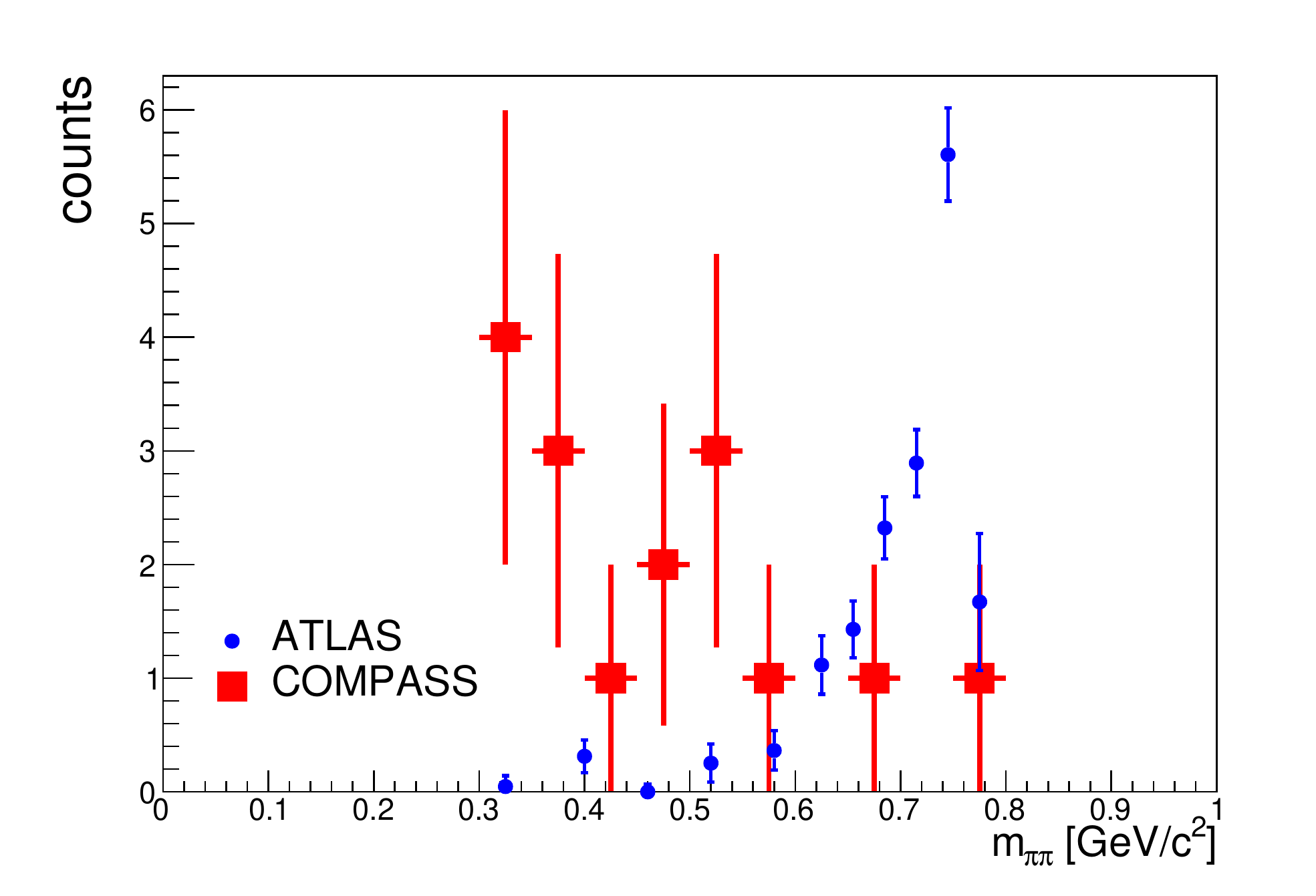}  
     \caption{\label{fig:CA}Invariant mass spectra for the $\pi^+\pi^-$ subsystem from the decay of X(3872) measured by COMPASS (red squares) and by ATLAS (blue points).}
\end{minipage}
 \end{figure}

Photoproduction of the $X(3872)$ at COMPASS is observed in the exclusive reaction $\mu^+ N \rightarrow \mu^+ N' X(3872)\pi^{\pm}\rightarrow \mu^+ N' J/\psi \pi^+\pi^-\pi^{\pm}$. The invariant mass spectrum of the $J\psi\pi^+\pi^-$ subsystem is shown in Fig. \ref{fig:x3872} (two entries per event). It demonstrates two peaks whose positions and widths 
 are as to be expected for the $\psi(2S)$ and $X(3872)$ states. The statistical significance of the $X(3872)$ signal depends on the applied selection criteria and varies from 4.5$\sigma$ to 6$\sigma$. The  cross section of the reaction $\gamma N \rightarrow N' X(3872)\pi^{\pm}$ multiplied by the branching fraction for the decay $X(3872)\rightarrow J/\psi\pi^+\pi^-$ was found to be $71\pm28_{stat}\pm39_{syst}$ pb in the covered kinematic range with a mean value of $\sqrt{s_{\gamma N}}=14$ GeV. It was also found that the shape of the invariant mass distribution for $\pi^+\pi^-$  produced from the $X(3872)$ decay (see Fig. \ref{fig:CA}) looks very different from previous results obtained by Belle, CDF, CMS and ATLAS. It could be an indication that the $X(3872)$ peak could contain a component with quantum numbers different from 1$^{++}$. 

The reaction with neutral exchange $\mu^+ N \rightarrow \mu^+ N' X(3872)\rightarrow \mu^+ N' J/\psi \pi^+\pi^-$ was investigated in parallel. No statistically significant evidence for an X(3872) signal was found. An upper limit for the production rate of $X(3872)$ in the process $\gamma N \rightarrow N' X(3872)$ was estimated to be 2.9 pb (CL=90\%).

More detailed information can be found in Ref. \cite{x3872}. 

\section{Photoproduction of $Z^{\pm}_c(3900)$}
\begin{figure}
 \begin{minipage}{18pc}
   \includegraphics[width=200px]{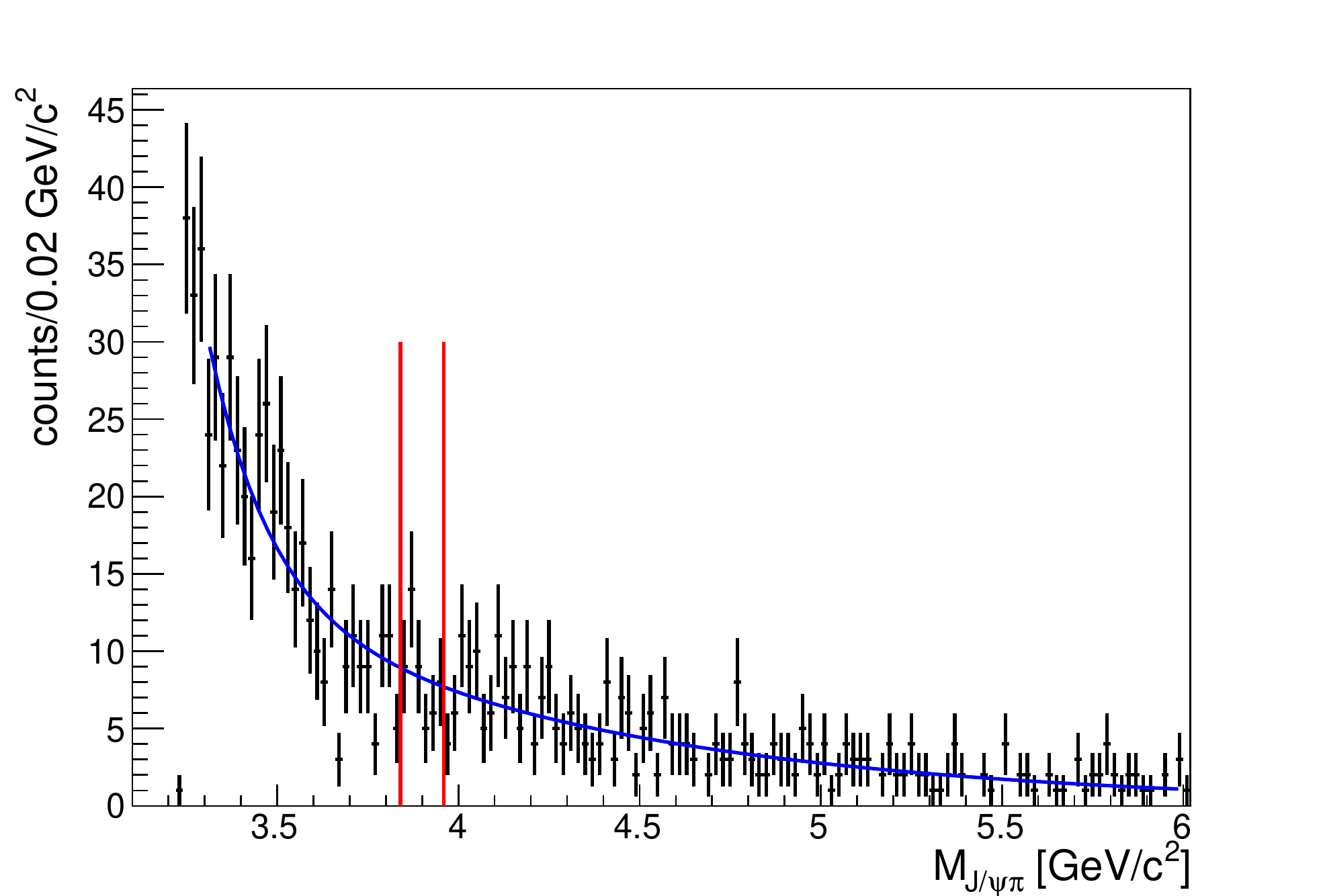}  
     \caption{\label{fig:zc}
The mass spectrum of the $J\psi\pi^{\pm}$ subsystem \cite{Zc}. The searching range is shown by the vertical lines while the curve represents the background fitting. }
\end{minipage}\hspace{2pc}%
\begin{minipage}{18pc}
   \includegraphics[width=200px]{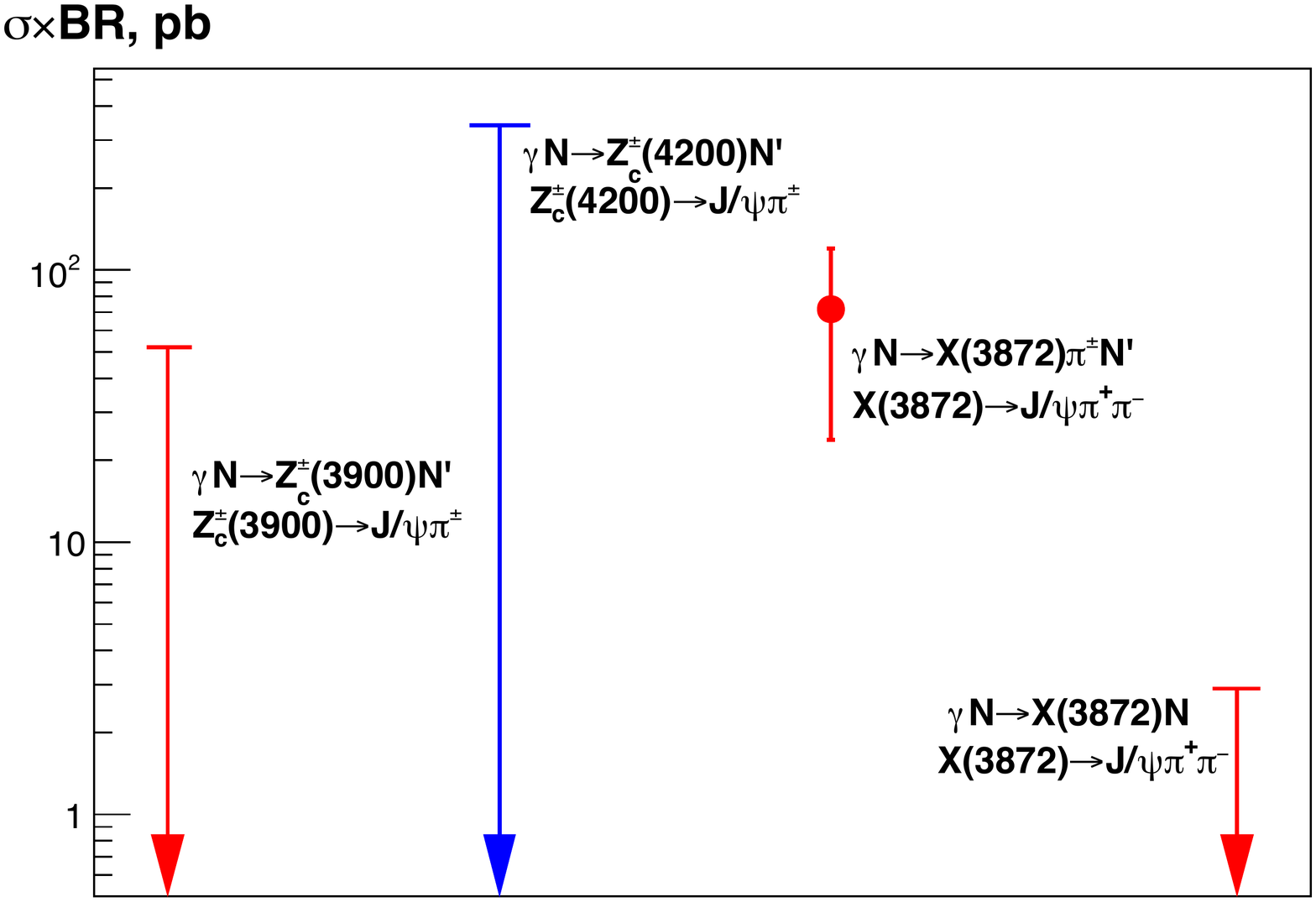}  
     \caption{\label{fig:all}COMPASS \cite{x3872, Zc} and COMPASS-based  \cite{Z4200} results for absolute rate (cross section multiplied by relative probability of corresponding decay) of photoproduction of $X(3872)$  and $Z_c(3900)$.}
\end{minipage}
 \end{figure}

The $Z^{\pm}_{c}(3900)$ state discovered via its decay into $J/\psi\pi^{\pm}$ by BESIII \cite{ZcBES} and Belle \cite{ZcBelle}  is one of the most promising tetraquark candidate. At COMPASS the search for $Z^{\pm}_{c}(3900)$  was performed in the exclusive reaction 
$\mu^+ N \rightarrow \mu^+ N' Z_{c}^{\pm}(3900)\rightarrow \mu^+ N' J/\psi \pi^{\pm}$. The $J/\psi\pi^{\pm}$ mass spectrum for exclusive events (see Fig. \ref{fig:zc}) does not exhibit any statistically significant resonant structure around the nominal mass of the $Z^{\pm}_{c}(3900)$.
Therefore an upper limit was determined for the product of the cross section of the $\gamma N \rightarrow N' Z_c^{\pm}(3900)$ process and the relative $Z^{\pm}_{c}(3900)\rightarrow J/\psi\pi^{\pm}$ decay probability to be 52 pb (CL=90\%) \cite{Zc}. An upper limit for the partial width of the $Z^{\pm}_{c}(3900)\rightarrow J/\psi\pi^{\pm}$ decay was also established basing on the production model described in \cite{Zcmod}.

The $J/\psi\pi^{\pm}$ mass spectrum measured by COMPASS was used in \cite{Z4200} to estimate the production rate of the $Z^{\pm}_c(4200)$, another exotic state also observed by Belle \cite{Z4200Belle}.

All COMPASS and COMPASS-based results for absolute rate (cross section multiplied by relative probability of corresponding decay) of photoproduction of exotic states are presented in Fig. \ref{fig:all}.

 
\section{New possibilities}
The upgrade of the COMPASS setup related with the data taking in 2016--2017 within the framework of the GPD program \cite{proposal2} provides new opportunities to search for direct production of exotic charmonium-like states. A new, 2.5 m long liquid hydrogen target ($\sim$0.27$X_{0}$) is much more transparent for photons than the $^6$LiD and NH$_3$ targets used before. The 
target is surrounded by a 4 m long recoil proton detector which can be used to reconstruct and identify recoil protons via time-of-flight and energy loss measurements. The existing system of two electromagnetic calorimeters is extended by installation of the new large-aperture calorimeter. With the new calorimetry system one can expect much better selection of exclusive events. Searching for production of the neutral $Z_c^0(3900)$, discovered by BES-III \cite{Zc0}, decaying into $J/\psi\pi^0$ as well as $J/\psi\eta$ final states will be possible. The final states containing the $\chi_{c0,1,2}$ -mesons could also be studied.

In more detail the question of the study of exotic charmonia at COMPASS is discussed in Ref. \cite{XYZmy}.

\section{Conclusions}
Photo(lepto-)production of exotic charmonia is a new promising production mechanism of the XYZ states tested by COMPASS. The $X(3872)$ meson became the first exotic charmonium-like state observed in photoproduction. The search results for exclusive production of the $Z_c^{\pm}(3900)$ state in the charge-exchange reaction are also reported. Despite a possibility to use data collected in 2016--2017 for the GPD program, statistics available at COMPASS for XYZ physics is limited. High-precision studies are foreseen at facilities with high-intensity photon beams like CLAS or GlueX.


\begin{thebibliography}{99}
\bibitem{photo0} Bing An Li,  Phys. Lett. \textbf{B605} (2005) 306.
\bibitem{photo1} Liu X.-H. Qiang Zhao, Frank E. Close, Phys. Rev. \textbf{D77} (2008) 094005.
\bibitem{photo2} He J., Liu X., Phys. Rev. \textbf{D80} (2009) 114007.
\bibitem {proposal} F. Bradamante, S Paul et al., CERN Proposal COMPASS,
      http://wwwcompass.cern.ch,\\ CERN/SPSLC 96-14, SPSC/P297; CERN/SPSLC 96-30, SPSC/P297,
     Addendum 1.
\bibitem{COMPASS1} COMPASS Collaboration, C. Adolph et al., 	Nucl. Instrum. Meth \textbf{A577} (2007) 455.
\bibitem{x3872belle}Belle Collaboration, S. K. Choi, et al.,  Phys. Rev. Lett. \textbf{91} (2003) 262001.
\bibitem{x3872LHCB1} LHCb Collaboration, R. Aaij, et al., Phys. Rev. Lett. \textbf{110}  (2013) 222001.
\bibitem{x3872LHCB2} LHCb Collaboration, R. Aaij et al., Phys. Rev. \textbf{D92} (2015) 011102.
\bibitem{x3872} COMPASS Collaboration, C. Adolph et al., CERN-EP-2017-165,  arXiv:1707.01796 (2017).
\bibitem{ZcBES} BESIII Collaboration M. Ablikim, et al., Phys. Rev. Lett. \textbf{110} (2013) 252001.
\bibitem{ZcBelle} Belle Collaboration, Z. Q. Liu, et al., Phys. Rev. Lett. \textbf{110}  (2013) 252002.
\bibitem{Zc} COMPASS Collaboration, C. Adolph et al.,  Phys. Lett. \textbf{B742} (2015) 330.
\bibitem{Zcmod} Q.-Y. Lin et al., Phys. Rev. \textbf{D88},  (2013) 114009.
\bibitem{Z4200}X.-Y. Wang, X.-R. Chen, A. Guskov, Phys. Rev. \textbf{D92} (2015) 094017.
\bibitem{Z4200Belle}Belle Collaboration, K. Chilikin et al., Phys. Rev. \textbf{D90}  (2014) 112009.
\bibitem{proposal2} COMPASS collaboration, COMPASS-II proposal, CERN/SPSC 2010-014, SPSC-P-340,     
\mbox{http://wwwcompass.cern.ch/compass/proposal/compass-II\_proposal/compass-II\_proposal.pdf}.
\bibitem{Zc0}BESIII Collaboration, M. Ablikim et al., Phys. Rev. Lett. \textbf{115} 112003.
\bibitem{XYZmy} I. Denisenko, A. Guskov, E. Mitrofanov, Phys. Part. Nucl. \textbf{48} (2017)  635. 
\end{thebibliography}
\end{document}